\documentclass[twocolumn,showpacs,prl,aps,floatfix,reprint]{revtex4-1}
\usepackage{amssymb}
\usepackage{amsmath}
\usepackage{graphicx}
\usepackage{dcolumn}
\usepackage{amsfonts}
\usepackage{bm}
\usepackage[dvipsnames]{color}
\newcommand{\bra}[1]{\left\langle #1\right|}
\newcommand{\ket}[1]{\left| #1\right\rangle}
\begin{document}

\title{Elastic nucleon-nucleus scattering as a direct probe of correlations beyond the independent-particle model}
\author{H. Dussan$^{1}$}
\author{M. H. Mahzoon$^{1}$}
\author{R. J. Charity$^{2}$}
\author{W. H. Dickhoff$^{1}$}
\author{A. Polls$^{3}$}
\affiliation{Department of Physics$^{1}$ and Chemistry$^{2}$, Washington University, St.
Louis, Missouri 63130}
\affiliation{ Departament d'Estructura i Constituents de la Mat\`{e}ria and Institut de Ci\`{e}nces del Cosmos$^{3}$, Universitat de Barcelona, Avda. Diagonal 647, E-8028 Barcelona, Spain}

\date{\today}

\begin{abstract}
Employing a recently-developed dispersive optical model (DOM) which allows a complete description of experimental data both above (up to 200 MeV) and below the Fermi energy in $^{40}$Ca, we demonstrate that elastic nucleon-nucleus scattering data constrain the spectral strength in the continuum of orbits that are nominally bound in the independent-particle model.
In the energy domain between 0 and 200 MeV, the integrated strength or depletion number is highly sensitive to the separation of the IPM orbit to the scattering continuum.
This sensitivity is determined by the influence of the surface-absorption properties of the DOM self-energy.
For an \textit{ab initio} calculation employing the self-energy of the charge-dependent Bonn (CDBonn) interaction which only includes the effect of short-range correlations, no such sensitivity is obtained and a depletion of 4\% is predicted between 0 and 200 MeV irrespective of the orbit.
The \textit{ab initio} spectral strength generated with the CDBonn interaction approaches the empirical DOM spectral strength at 200 MeV.
Both spectral distributions allow for an additional 3-5\% of the strength at even higher energies which is associated with the influence of short-range correlations.
We suggest that the non-local form of the DOM allows for an analysis of elastic-nucleon-scattering data
that directly determines the depletion of bound orbits.
While obviously relevant for the analysis of elastic nucleon scattering on stable targets, this conclusion holds equally well for experiments involving rare isotopes in inverse kinematics as well as experiments with electrons on atoms or molecules. 
\end{abstract}

\maketitle

An accurate determination of the limits of the nuclear independent-particle model (IPM) represents an important goal in the study of exotic nuclei that can be probed at rare-isotope facilities.
To retrieve this information, we mostly rely on strongly-interacting probes and therefore it is more important than ever to strengthen the link between the descriptions of nuclear reactions and nuclear structure~\cite{Dickhoff10a}.
An important candidate for this link is provided by the dispersive optical model (DOM) pioneered by Mahaux and Sartor~\cite{Mahaux91} which considers the potentials a nucleon experiences above and below the Fermi energy as representing different aspects of one underlying self-energy.
The DOM also provides an ideal strategy to predict properties of exotic nuclei by utilizing extrapolations of these potentials towards the respective drip lines~\cite{Charity06,Charity07,Mueller11,Waldecker2011}.
It was recently demonstrated in Ref.~\cite{Hossein14} that a nonlocal representation of the self-energy
of ${}^{40}$Ca is essential to describe \textit{all} available data below the Fermi energy that can be linked to the nucleon  propagator~\cite{Dickhoff08} while maintaining a correct description of elastic-scattering data. 
The resulting DOM potential represents the nucleon self-energy constrained by all available experimental data up to 200 MeV.
This self-energy allows for a consistent treatment of nuclear reactions that require distorted waves generated by optical potentials as well as overlap functions and their normalization for the addition and removal of nucleons to discrete final states. 
 
In the present work we demonstrate that the DOM is capable of identifying the presence of spectral strength in the continuum for predominantly bound orbits thereby providing a quantitative description of their depletion.
Such an analysis can in principle also be applied to other finite systems, for example for elastic electron scattering on atoms or molecules.
The depletion of the orbits near the Fermi energy has been studied with the $(e,e'p)$ reaction~\cite{Herder88,Sick91,Lapikas93,Pandharipande97} suggesting spectroscopic factors that are 35 to 40\% lower than IPM values~\cite{Lapikas93}.
Theoretical work reviewed in Ref.~\cite{Dickhoff04} assigns about one third of the reduction of the valence strength to the influence of short-range correlations (SRC) that both deplete the IPM Fermi sea as well as  generate the observed 10\% of high-momentum nucleons in the ground state~\cite{Rohe04}. 
The remaining reduction should reflect the coupling to low-lying surface excitations representing long-range correlations (LRC). 

Spectroscopic factors in nuclear physics have a long history~\cite{Macfarlane60} and their strict observability remains a topic of debate~\cite{Furnstahl02,Mukh10}.
The importance of relative spectroscopic factors is well established~\cite{Kramer01,Schiffer12}.
Given a Hamiltonian, spectroscopic factors can be unambiguously defined using the Green's function method as the normalization of the overlap function for the removal or addition of a nucleon from or to the ground state of an even-even nucleus ending in a bound state.
The overlap function solves the Dyson equation at the energy corresponding to the state associated with the removed or added particle~\cite{Dickhoff08}.
Since all many-body methods applied to nuclei start from a Hamiltonian for nucleons, it seems to be a reasonable assumption that the notion of a normalized overlap function will remain a useful concept.

The extraction of spectroscopic factors from the $(e,e'p)$ reaction is by no means a closed subject as a different treatment of the outgoing-proton distorted waves~\cite{Udias95} yields numbers that are about 0.10 to 0.15 larger than those extracted by the NIKHEF group~\cite{Lapikas93}. 
The DOM analysis for ${}^{40}$Ca~\cite{Hossein14} also generates values that are about 0.15 larger than those deduced by  the NIKHEF group~\cite{Kramer89} suggesting that a re-analysis of these data with DOM ingredients is pertinent.

Complementary information is provided by the corresponding strength above the Fermi energy which lies in the scattering continuum.
This information is contained in the so-called particle spectral function which has not been generated for a finite system with the inclusion of SRC up to now, but has been studied for infinite nuclear matter~\cite{Ramos89,Vonderfecht91,Benhar92,Vonderfecht93}.
Some theoretical works address the influence of LRC on the particle spectral function but do not treat the continuum aspects~\cite{Barbieri1,Barbieri2,Barbieri09,Waldecker2011}.
In the DOM, it is constrained by experimental data from elastic scattering of neutrons and protons from 0 to 200 MeV for ${}^{40}$Ca.
Further insight is provided by contrasting the DOM particle spectral function with the one calculated from the CDBonn interaction~\cite{Machleidt1995} using the method developed for ${}^{16}$O~\cite{Muther94,Muther95} and recently applied in Ref.~\cite{Dussan11} to ${}^{40}$Ca which emphasizes the role of SRC.

The particle spectral function for a finite system requires the complete reducible self-energy $\Sigma$.
We have employed a momentum-space scattering code~\cite{Dussan11} to calculate $\Sigma$.
In an angular-momentum basis, iterating the irreducible self-energy $\Sigma^*$ to all orders, yields
\begin{eqnarray}\label{eq:redSigma1}
\Sigma_{\ell j}(k,k^\prime ;E)& = &\Sigma^*_{\ell j}(k,k^\prime ;E)\\ \nonumber
  &+&  \!\!       \int \!\! dq q^2 \Sigma^*_{\ell j}(k,q;E)G^{(0)}(q;E )\Sigma_{\ell j}(q,k^\prime ;E) ,
\end{eqnarray}
where $G^{(0)}(q; E ) = (E - \hbar^2q^2/2m + i\eta)^{-1}$ is the free propagator.
The propagator is then obtained from the Dyson equation in the following form~\cite{Dickhoff08}
\begin{eqnarray}
G_{\ell j}(k, k^{\prime}; E) &=& \frac{\delta( k - k^{\prime})}{k^2}G^{(0)}(k; E)  
 \label{eq:gdys1}  \\
		&+&		      G^{(0)}(k; E)\Sigma_{\ell j}(k, k^{\prime}; E)G^{(0)}(k; E) .
	\nonumber	
\end{eqnarray}
The on-shell matrix elements of the reducible self-energy in Eq.~(\ref{eq:redSigma1}) are sufficient to describe all aspects of elastic scattering like differential, reaction, and total cross sections as well as polarization data~\cite{Dussan11}.
While only this element is strictly related to data, the DOM analysis in its nonlocal implementation correctly and simultaneously describes all relevant data at all energies (up to 200 MeV) both below and above the Fermi energy~\cite{Hossein14}.
We therefore proceed on the assumption that the DOM reducible self-energy is reasonably unique and may therefore provide insight into the depletion of the Fermi sea.
The spectral representation of 
the particle part of the propagator, referring to the $A+1$ system, appropriate for a treatment of the continuum and possible open channels is given by~\cite{Mahaux91}
\begin{eqnarray}
G_{\ell j}^{p}(k ,k' ; E)  &=&  
\sum_n  \frac{ \phi^{n+}_{\ell j}(k) \left[\phi^{n+}_{\ell j}(k')\right]^*
}{ E - E^{*A+1}_n +i\eta }   \label{eq:propp} \\
& + & 
\sum_c \int_{T_c}^{\infty} dE'\  \frac{\chi^{cE'}_{\ell j}(k) \left[\chi^{cE'}_{\ell j}(k')\right]^* }{
E - E' +i\eta} . 
\nonumber
\end{eqnarray}
Overlap functions for bound $A+1$ states are given by $ \phi^{n+}_{\ell j}(k)=\bra{\Psi^A_0} a_{k\ell j}
\ket{\Psi^{A+1}_n}$, whereas those in the continuum are given by $ \chi^{cE}_{\ell j}(k)=\bra{\Psi^A_0} a_{k\ell j} \ket{\Psi^{A+1}_{cE}}$ indicating the relevant channel by $c$ and the energy by $E$.
Excitation energies in the $A+1$ system are with respect to the $A$-body ground state $E^{*A+1}_n = E^{A+1}_n -E^A_0$.
Each channel $c$ has an appropriate threshold indicated by $T_c$ which is the experimental threshold with respect to the ground-state energy of the $A$-body system.
The overlap function for the elastic channel can be explicitly calculated by solving the Dyson equation while it is also possible to obtain the complete spectral density for $E>0$ 
\begin{eqnarray}
S_{\ell j}^{p}(k ,k' ; E) 
=
\sum_c \chi^{cE}_{\ell j}(k) \left[ \chi^{cE}_{\ell j}(k') \right]^* .
\label{eq:specp}
\end{eqnarray}
In practice, this requires solving the scattering problem twice at each energy so that one may employ
\begin{eqnarray}
\!\! S_{\ell j}^{p}(k ,k' ; E) 
= \frac{i}{2\pi} \left[ G_{\ell j}^{p}(k ,k' ; E^+) - G_{\ell j}^{p}(k ,k' ; E^-) \right]
\label{eq:specpp}
\end{eqnarray}
with $E^\pm =E\pm i\eta$, and only the elastic-channel contribution to Eq.~(\ref{eq:specp}) is explicitly known.
Equivalent expressions pertain to the hole part of the propagator $G_{\ell j}^{h}$~\cite{Mahaux91}.

\begin{figure}[tbp]
\includegraphics*[scale=0.38]{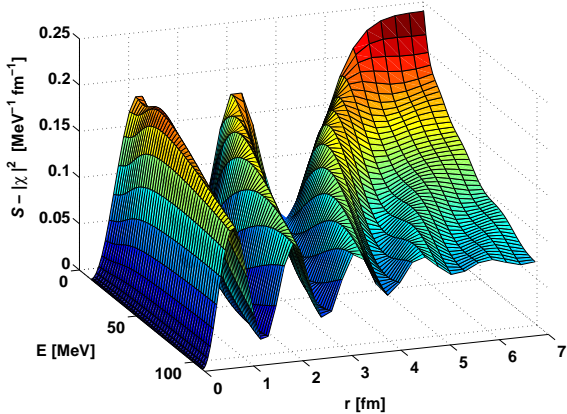}
\caption{(Color online) Difference between the particle spectral function for $s_{1/2}$ and the contribution of the elastic-scattering wave function multiplied by $r^2$, as a function of both energy and position. Asymptotically with $r$, this difference is approximately constant and determined only by the inelasticity.}
\label{fig:Sandpsi}
\end{figure}

The calculations are performed in momentum space according to Eq.~(\ref{eq:redSigma1}) to generate the off-shell reducible self-energy and thus the spectral density by employing Eqs.~(\ref{eq:gdys1}) and (\ref{eq:specpp}).
Because the momentum-space spectral density contains a delta-function associated with the free propagator, it is convenient for visualization purposes to consider a Fourier transform to coordinate space 
\begin{eqnarray}
S_{\ell j}^{p}(r ,r' ; E) &=& \frac{2}{\pi} \label{eq:specpr} \\
& \times &  \int \!\! dk k^2 \! \int \!\! dk' k'^2 j_\ell(kr) S_{\ell j}^{p}(k ,k' ; E) j_\ell(k'r') ,
\nonumber 
\end{eqnarray}
which has the physical interpretation for $r=r'$ as the probability density $S_{\ell j}(r;E)$ for adding a nucleon with energy $E$ at a distance $r$ from the origin for a given $\ell j$ combination.
By employing the asymptotic analysis to the propagator in coordinate space following \textit{e.g.} Ref.~\cite{Dickhoff08}, one may express the elastic-scattering wave function that contributes to Eq.~(\ref{eq:specp}) in terms of the half on-shell reducible self-energy obtained according to
\begin{eqnarray}
\chi^{el E}_{\ell j}(r) & = &\left[ \frac{2mk_0}{\pi \hbar^2} \right]^{1/2} \bigg\{ j_\ell(k_0r)  \label{eq:elwf} \\
& + & \left. \int \!\! dk k^2 j_\ell(kr) G^{(0)}(k;E) \Sigma_{\ell j}(k,k_0;E) \right\} ,
\nonumber
\end{eqnarray}
where $k_0$ is related to the scattering energy in the usual way.
We subtract this contribution to the spectral function given by its absolute square from $S_{\ell j}(r;E)$ in Fig.~\ref{fig:Sandpsi} for different energies.
Asymptotically at large distances, the influence of other open channels is represented by an almost constant shift whereas, inside the range of the potential, a pattern related to the absorptive properties of the potential and the orbits that are occupied emerges.

The presence of strength in the continuum associated with mostly-occupied orbits (or mostly empty but $E<0$  orbits) is obtained by double folding the spectral density in Eq.~(\ref{eq:specpr}) in the following way
\begin{eqnarray}
\!\!\! S_{\ell j}^{n+}(E) 
=  \int \!\! dr r^2 \!\! \int \!\! dr' r'^2 \phi^{n-}_{\ell j}(r) S_{\ell j}^{p}(r ,r' ; E) \phi^{n-}_{\ell j}(r') ,
\label{eq:specfunc}
\end{eqnarray}
using an overlap function 
\begin{equation}
\sqrt{S^n_{\ell j}} \phi^{n-}_{\ell j}(r)=\bra{\Psi^{A-1}_n} a_{r\ell j} \ket{\Psi^{A}_0} , 
\label{eq:overm}
\end{equation}
corresponding to a bound orbit with $S^n_{\ell j}$ the relevant spectroscopic factor and $\phi^{n-}_{\ell j}(r)$  normalized to 1~\cite{Dussan11}.

In the case of an orbit below the Fermi energy, this strength identifies where the depleted strength resides in the continuum.
The occupation number of this orbit is given by an integral over a corresponding folding of the hole spectral density
\begin{eqnarray}
\!\!\!\!\!\!\! S_{\ell j}^{n-}(E) 
= \!\! \int \!\! dr r^2 \!\! \int \!\! dr' r'^2 \phi^{n-}_{\ell j}(r) S_{\ell j}^{h}(r ,r' ; E) \phi^{n-}_{\ell j}(r') ,
\label{eq:spechr}
\end{eqnarray}
where $S_{\ell j}^{h}(r,r';E)$ provides equivalent information below the Fermi energy as $S_{\ell j}^{p}(r,r';E)$ above.
An important sum rule is valid for the sum of the occupation number for the orbit $n_{n \ell j}$ and its depletion number $d_{n \ell j}$~\cite{Dickhoff08}
\begin{eqnarray}
\!\! 1 =  n_{n \ell j} + d_{n \ell j} \!\!
\label{eq:sumr} 
 =\!\! \int_{-\infty}^{\varepsilon_F} \!\!\!\! dE\ S_{\ell j}^{n-}(E) \!+ \!\! \int_{\varepsilon_F}^{\infty} \!\!\!\! dE\ S_{\ell j}^{n-}(E)  ,
\end{eqnarray}
equivalent to $a^\dagger_{n \ell j} a_{n \ell j} +a_{n \ell j}a^\dagger_{n \ell j} =1$.
The average Fermi energy $\varepsilon_F \equiv \frac{1}{2} \left[ (E^{A+1}_0-E^A_0) + (E^A_0 - E^{A-1}_0) \right]$  is introduced here~\cite{Mahaux91}.

Strength above $\varepsilon_F$, as expressed by Eq.~(\ref{eq:specfunc}), reflects the presence of the imaginary self-energy at positive energies.
Without it, the only contribution to the spectral function comes from the elastic channel.
The folding in Eq.~(\ref{eq:specfunc}) then involves integrals of orthogonal wave functions and yields zero.
Because it is essential to describe elastic scattering with an imaginary potential, 
it automatically ensures that the elastic channel does not exhaust the spectral density and therefore some spectral strength associated with IPM bound orbits also occurs in the continuum.

We also present results for a microscopic calculation of the ${}^{40}$Ca self-energy obtained from the CDBonn interaction~\cite{Machleidt1995}.
Details have been provided in Ref.~\cite{Dussan11}.
Because all ingredients of this calculation involve momentum-space quantities, the double folding in Eq.~(\ref{eq:specfunc}) is performed in momentum space utilizing overlap functions obtained in Ref.~\cite{Dussan11}. 
The experimentally-constrained nonlocal DOM potential of Ref.~\cite{Hossein14} was Fourier transformed to momentum space to allow the calculation of the off-shell reducible self-energy of Eq.~(\ref{eq:redSigma1}).
We have therefore employed the neutron self-energy for this calculation but the proton self-energy is identical apart from the Coulomb term.
Fourier transforming the spectral density according to Eq.~(\ref{eq:specpr}) allows further analysis and also to perform the folding with the bound DOM overlap functions obtained in coordinate space~\cite{Dickhoff10}.

\begin{figure}[tbp]
\includegraphics*[scale=0.32]{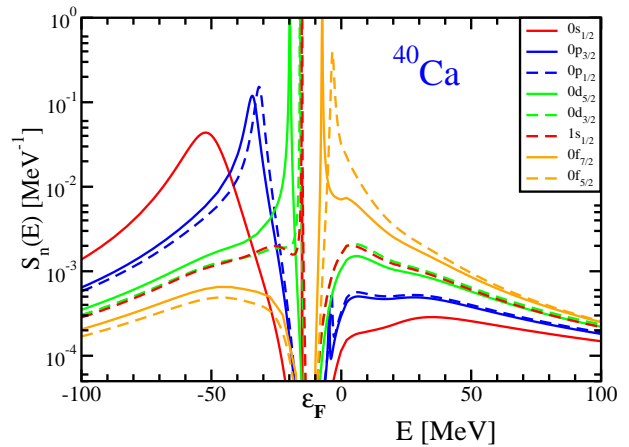}
\caption{(Color online) Calculated spectral strength, both below and above the Fermi energy, for bound orbits in ${}^{40}$Ca. The spectral strength is constrained by elastic-scattering data, level structure, charge density, particle number, and the presence of high-momenta below the Fermi energy~\protect\cite{Hossein14}.}
\label{fig:deplE}
\end{figure}

We display in Fig.~\ref{fig:deplE} the results of the DOM spectral function for the most relevant bound orbits in ${}^{40}$Ca including the hole spectral function of Ref.~\cite{Hossein14}.
Because the DOM analysis assumes that the imaginary part of the self-energy starts at $\varepsilon_F$, the  spectral strength is a continuous function of the energy.
The method of solving the Dyson equation for $E<0$ is very different than that for $E>0$.
The continuity of the curves at $E=0$ confirms the numerical aspects of both of these calculations. 
Below the Fermi energy, the spectral strength contains peaks associated with the $0s_{1/2},0p_{3/2},0p_{1/2},0d_{5/2},1s_{1/2},$ and $0d_{3/2}$ orbits with narrower peaks for orbits closer to the Fermi energy.
Their strength was calculated for the overlap functions associated with the location of the peaks by solving the Dyson equation without the imaginary part but with self-consistency for the energy of the real part~\cite{Dickhoff10}.
The strength of these orbits above the Fermi energy exhibits systematic features displaying more strength when the IPM energy is closer to the continuum threshold.
We make this observation quantitative by listing the integrated strength according to the terms of Eq.~(\ref{eq:sumr}) in Table~\ref{Tbl:depln}. 
For the depletion we integrate from 0 to 200 MeV which corresponds to the energy domain constrained by data in the DOM.
\begin{table}[tbp]
\caption{Occupation and depletion numbers for bound orbits in ${}^{40}$Ca. The $d_{nlj}[0,200]$ depletion numbers have only been integrated from 0 to 200 MeV. The fraction of the sum rule in Eq.~(\ref{eq:sumr}) that is exhausted, is illustrated by $n_{n \ell j} + d_{n \ell j}[\varepsilon_F,200]$. We also list the $d_{nlj}[0,200]$ depletion numbers for the CDBonn calculation in the last column.}
\label{Tbl:depln}%
\begin{ruledtabular}
\begin{tabular}{crrcl}
orbit  & $n_{n \ell j}$ & $d_{n \ell j}[0,200]$ & $n_{n \ell j} + d_{n \ell j}[\varepsilon_F,200]$ & $d_{n_\ell j}[0,200]$ \\
& DOM & DOM & DOM & CDBonn \\
\hline
$0s_{1/2}$ & 0.926 & 0.032  & 0.958 & 0.035\\
$0p_{3/2}$& 0.914 & 0.047 & 0.961 & 0.036 \\
$1p_{1/2}$ &  0.906 & 0.051 &0.957 & 0.038 \\
$0d_{5/2}$ & 0.883 & 0.081 & 0.964 & 0.040 \\
$1s_{1/2}$ & 0.871 & 0.091 & 0.962 & 0.038 \\ 
$0d_{3/2}$ & 0.859 & 0.097 & 0.966 & 0.041 \\
$0f_{7/2}$ & 0.046 &  0.202 & 0.970 & 0.034 \\
$0f_{5/2}$ & 0.036  & 0.320 & 0.947 & 0.036 \\
\end{tabular}
\end{ruledtabular}
\end{table}
We also include the $0f_{7/2}$ and $0f_{5/2}$ spectral functions in Fig.~\ref{fig:deplE} and corresponding results in Table~\ref{Tbl:depln} noting that the strength in the continuum from 0 to 200 MeV further rises to 0.202 and 0.320, respectively. 
From $\varepsilon_F$ to 0 the strength for these states is also included in the sum and decreases from 0.722 to 0.591, respectively.
This illustrates that there is a dramatic increase of strength into the continuum when the IPM energy approaches this threshold.
Such orbits correspond to valence states in exotic nuclei~\cite{Gade04,Jensen11,Charity14}.
The $1p_{3/2}$ and $1p_{1/2}$ spectral functions are not shown as they mimic the behavior of the $0f_{7/2}$ distribution but their presence causes the wiggles in the $0p_{3/2}$ and $0p_{1/2}$ spectral functions due to slight nonorthogonality.

\begin{figure}[tbp]
\includegraphics*[scale=0.3]{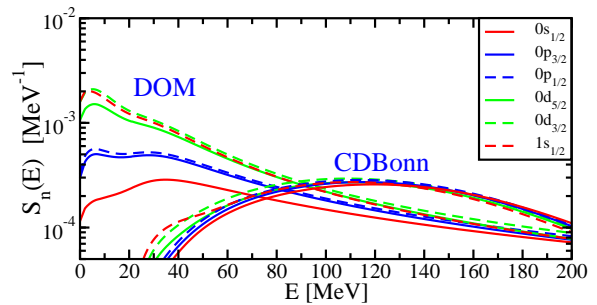}
\caption{(Color online) Calculated spectral strength for mostly occupied orbits in ${}^{40}$Ca from 0 to 200 MeV.
The CDBonn spectral functions exhibit mainly volume absorption.}
\label{fig:deplEX}
\end{figure}

This sensitivity to the separation from the continuum is associated with the pronounced surface absorption necessary to describe the elastic-scattering data in this energy range. 
At higher energies, volume absorption dominates and the strengths of the different orbits become similar as illustrated in Fig.~\ref{fig:deplEX}.
This figure also includes the CDBonn predictions which highlight the notion that SRC predominantly impact higher energies. 
While the CDBonn spectral function overestimates the DOM results above 100 MeV, it is quite likely that a somewhat harder interaction like Argonne V18~\cite{Wiringa1995} would move some of this excess strength to higher energy~\cite{Muther05}.

The fraction of the sum rule of Eq.~(\ref{eq:sumr}) for the DOM in Table~\ref{Tbl:depln} indicates that a few percent  of the strength occurs at energies higher than 200 MeV.
Theoretical work associates such strength with SRC~\cite{Vonderfecht91}.
No surface absorption is present in the microscopic CDBonn results and their depletions in Table~\ref{Tbl:depln} correspond to a uniform strength distribution for all orbits consistent with the SRC interpretation as illustrated in Fig.~\ref{fig:deplEX}.
As for the DOM, about 3-5\% of the strength occurs above 200 MeV when the occupation numbers of Ref.~\cite{Dussan11} are included. 

Our analysis clarifies that elastic-scattering data, combined with a complete treatment of sp properties below the Fermi energy, provide a quantitative demonstration for the presence of some continuum strength for predominately bound states.
This conclusion is possible because scattering and structure data are analyzed using the full nonlocal treatment of the DOM~\cite{Hossein14} which guarantees the proper treatment of sum rules like Eq.~(\ref{eq:sumr}).
The location of the strength closely tracks the absorptive properties of the self-energy and exhibits a pronounced dependence on the separation of the IPM energy to the continuum.
An \textit{ab initio} calculation of the nucleon self-energy, based on the CDBonn potential which only treats SRC,  generates a modest depletion without any pronounced dependence on the location of the orbit.
Both the CDBonn calculation and the DOM analysis allow for a few percent of strength beyond 200 MeV.
Our results therefore illustrate the importance of measuring elastic-nucleon-scattering data for exotic nuclei in inverse kinematics.
A nonlocal DOM analysis can then directly assess how correlations for nucleons change when the drip lines are approached.
Analysis of transfer reactions with ingredients of the nonlocal DOM treatment may shed further quantitative light on neutron properties in such systems~\cite{Nguyen2011,Timofeyuk13,Titus14}.



This material is based upon work supported by the U.S. Department of Energy, Office of Science, Office of Nuclear Physics under Award number DE-FG02-87ER-40316, by the U.S. National Science Foundation under grant PHY-1304242, and Grant No. FIS2011-24154 from MICINN (Spain) and  Grant No. 2014SGR-401
from Generalitat  de Catalunya. One of us (W.H.D.) acknowledges the fruitful collaboration with H.~M\"{u}ther from the University of T\"{u}bingen, Germany, where some of the present work was initiated.

\bibliography{DOMbib_W}

\end{document}